\documentclass[aip,apl,preprint,superscriptaddress,showpacs,citeautoscript]{revtex4-1}

\usepackage{graphicx}
\usepackage{amsmath}
\usepackage{amssymb}
\usepackage{bm}
\usepackage{color}
\usepackage{array}
\usepackage{epstopdf}

\newcolumntype {s}[1]{@{\hspace{#1}}} % space

\graphicspath{{.}{./EPS/}}

\newcommand{\ket}[1]{\left | \, #1 \right \rangle}
\newcommand{\bra}[1]{\left \langle #1 \, \right |}

\newcommand*{\vek}[1]{{\bm{\mathrm{#1}}}}

\DeclareMathSymbol{\myRe}{\mathord}{symbols}{"3C}

\DeclareMathSymbol{\myIm}{\mathord}{symbols}{"3D}

\begin{document}

\title{Valley filter from magneto-tunneling between single and bi-layer graphene}

\author{L. Pratley}
%\email{luke.pratley@gmail.com}
\affiliation{School of Chemical and Physical Sciences and MacDiarmid
Institute for Advanced Materials and Nanotechnology, Victoria
University of Wellington, PO Box 600, Wellington 6140, New Zealand}

\author{U. Z\"ulicke}
\email{uli.zuelicke@vuw.ac.nz}
%\thanks{}
\affiliation{School of Chemical and Physical Sciences and MacDiarmid
Institute for Advanced Materials and Nanotechnology, Victoria
University of Wellington, PO Box 600, Wellington 6140, New Zealand}
\affiliation{Kavli Institute for Theoretical Physics, University of California,
Santa Barbara, CA 93106, USA}

\date{\today}

\begin{abstract}
We consider tunneling transport between two parallel graphene sheets
where one is a single-layer sample and the other one a bi-layer. In the
presence of an in-plane magnetic field, the interplay between combined
energy and momentum conservation in a tunneling event and the distinctive
chiral nature of charge carriers in the two systems turns out to favor tunneling
of electrons from one of the two valleys in the graphene Brillouin zone.
Adjusting the field strength enables manipulation of the valley polarization
of the current, which reaches its maximum value of 100\% concomitantly
with a maximum of the tunneling conductance.
\end{abstract}

%\keywords{}

\maketitle

\begin{figure}[t]
\includegraphics[width=0.5\columnwidth]{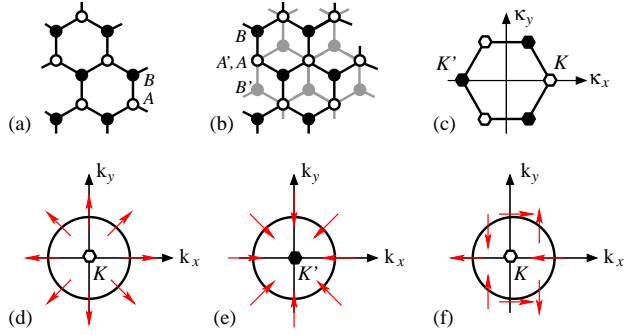}
\caption{\label{fig:basics}
Structural and electronic properties of graphene. (a)~Single-layer graphene
has carbon atoms arranged on a honeycomb lattice. Sites on its two equivalent
sublattices are indicated by empty and filled circles and labelled $A$ and $B$,
respectively. Electrons localized on the $A$ ($B$) sublattice are in pseudo-spin-up
(down) eigenstates. (b)~In bi-layer graphene, atoms from the $A$ and $A'$
sublattices of the two layers are bonded vertically, while those on $B$ and $B'$
sites are dangling. Electronic excitations at low energy are from the subspace
spanned by $B$ and $B'$ states, which correspond to the pseudo-spin
eigenstates for electrons in bi-layer graphene. (c)~The first Brillouin zone of
single-layer (and bi-layer) graphene. High-symmetry $\vek{K}$ ($\vek{K'}$)
points are indicated by empty (filled) hexagons. (d)--(f)~Pseudo-spin polarization
of conduction-band electrons (indicated by red arrows) for the $\vek{K}$ valley of
single-layer graphene [(d)], the $\vek{K'}$ valley of single-layer graphene [(e)],
and the $\vek{K}$ valley of bi-layer graphene [(f)].}
\end{figure}

%\textit{Introduction} --
The concept of \emph{spintronics\/}~\cite{zut04} continues to stimulate the
detailed study of intertwined dynamics of intrinsic (pseudo-)spin-1/2 degrees
of freedom and the orbital motion of charge carriers~\cite{sin12}.
Graphene-based nanomaterials~\cite{wei12} are particularly attractive
systems for spintronics applications~\cite{pes12} because, in addition to
their ordinary spin, electrons in graphene also carry an orbital pseudo-spin
and a valley-isospin degree of freedom~\cite{cas09}. While the
pseudo-spin-up/down eigenstates correspond to an electron's position on
the two equivalent sublattices of graphene's honeycomb lattice in real space,
the valley-isospin quantum number distinguishes states near the $\vek{K}$
and $\vek{K^\prime}\equiv -\vek{K}$ high-symmetry points in reciprocal space.
Schematics of the single-layer and bi-layer graphene lattices as well as their
(identical) Brillouin zone(s) are shown in panels (a)--(c) of Fig.~\ref{fig:basics}.

The possibility to realize valley-isospin-based spintronics , called
\emph{valleytronics\/}, in graphene has attracted a lot of
interest~\cite{xia07,ryc07,yao08,gar08,abe09,fuj10,low10,zha10,sch10,wu11,
gun11,gol11,jia13,wu13,kha13}. The operation of valleytronic devices generally
depends on the ability to generate valley-asymmetric charge currents.
Mechanisms to separately address electrons from individual valleys necessarily
involve the breaking of inversion and/or time-reversal symmetries~\cite{xia07},
e.g., via nanostructuring~\cite{ryc07,sch10}, coupling
to electromagnetic fields~\cite{yao08,gar08,abe09,fuj10,low10,zha10,gol11,wu13,kha13},
application of mechanical strain~\cite{fuj10,low10,zha10,wu11,jia13,kha13},
or presence of defects~\cite{gun11}. In many of these situations, the mobility of
charge carriers could be compromised by the required inhomogeneity and/or orbital
effects of the applied external fields.

Here we propose a valley-filter device that is based on resonant electron
tunneling between parallel single-layer and bi-layer graphene sheets. Vertical
heterostructures of two-dimensional (2D) crystals have recently been
fabricated~\cite{brit12,brit12a,geo12,brit13,gei13,myo13}, and their
electronic properties have become the subject of detailed theoretical
study~\cite{fee12,bal12,vas13,pra13}. As discussed below, application of a
magnetic field \emph{parallel\/} to the graphene sheets enables direct
tuning of the valley polarization of the tunneling current, with a possible
maximum value of 100\%. In contrast to other valley-filter designs, no significant
modification of the graphene sheets' electronic structure is required to
enable valley-asymmetric transport. As an additional feature, a
valley-polarized current is generated simultaneously in both the
single-layer and the bi-layer sheets.

Our proposed valley-filter design utilizes the strong link between linear orbital
momentum, pseudo-spin and valley isospin for electrons in graphene materials.
Due to peculiarities of the honeycomb-lattice band structure, an electron's
pseudo-spin state is locked to its linear motion on the 2D graphene
sheet in a way that is normally exhibited by ultra-relativistic (massless) Dirac
fermions~\cite{cas09}. The exact form of this coupling turns out to be different
for electrons from the two valleys and also depends on the type of graphene
structure, e.g., whether it is a single-layer or a bi-layer sample. See
Fig.~\ref{fig:basics}, panels (d)--(f), for an illustration. Furthermore, states with
the same wave-vector shift $\vek{k}\equiv (k_x, k_y)$ from the high-symmetry
point in the conduction and valence bands have opposite pseudo-spin polarization.
The characteristic differences between valley-dependent pseudo-spin-momentum
locking in single-layer and bi-layer graphene makes it possible to achieve valley
separation in a momentum-resolved tunneling structure proposed here.

%\textit{Tunneling in vertical graphene heterostructures} -- 
Tunneling transport between parallel 2D conductors exhibits
strongly resonant behavior~\cite{smo89a,eis91,hay91,gen91,sim93} as a function
of in-plane magnetic field $\vek{B}_\|$ and bias voltage $V$ because of the
requirement to simultaneously conserve the canonical in-plane momentum and
total energy of every tunneled electron. Theoretical descriptions of 2D-to-2D
tunneling~\cite{zhe93,lyo93,rai96,jun96} have been able to explain the experimental
observations with great accuracy.

\begin{figure}[t]
\includegraphics[width=0.42\columnwidth]{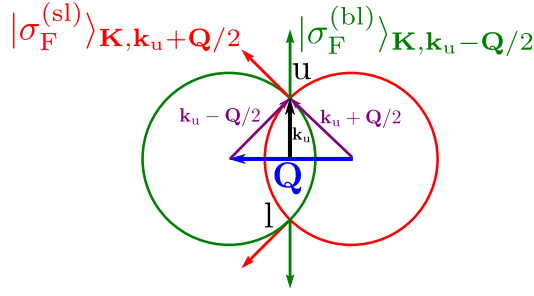}
\caption{\label{fig:pseudoSchem}
Visualization of momentum-resolved tunneling between 2D systems in the presence
of an in-plane magnetic field. The requirement of simultaneous energy and momentum
conservation restricts tunneling to states at intersection points of the Fermi surfaces
for the two layers. The pseudo-spin of these states is fixed by the kinetic momentum for
electrons in single-layer and bi-layer graphene, and the tunneling probability depends
on the matrix element given in Eq.~(\ref{eq:overlap}).}
\end{figure}

Recent progress in fabricating tunnel-coupled unconventional 2D systems where
the charge carriers' intrinsic (pseudo-)spin-1/2 degree of freedom is rigidly locked
to their orbital motion has stimulated further interest~\cite{per12,vas13,pra13}. In
particular, it has been shown~\cite{pra13} that the linear (i.e., small-bias)
magneto-tunneling conductance for electrons from valley $\gamma$ ($=\vek{K}$
or $\vek{K'}$ in graphene, or $\vek{\Gamma}$ in an ordinary 2D quantum-well
system) is given by 
\begin{eqnarray}\label{eq:conductance}
G^{(\gamma)} &=& \frac{g_{\text{s}} e^2}{\hbar}  2\pi A \,
\rho_{\text{F}}^{(1)}\, \rho_{\text{F}}^{(2)} \, \left[ \left|
\Gamma_{\text{u}}^{(\gamma)}(\vek{Q})\right|^2 + \left|
\Gamma_{\text{l}}^{(\gamma)}(\vek{Q})\right|^2 \right] 
\nonumber \\ && \hspace{5cm}
\times \frac{\Theta\left( | \vek{Q} | - \left| k_{\text{F}}^{(1)} -
k_{\text{F}}^{(2)} \right| \right)\Theta \left( k_{\text{F}}^{(1)} +
k_{\text{F}}^{(2)} - | \vek{Q} | \right)}{\sqrt{\left[\left( k_{\text{F}}^{(1)}
+ k_{\text{F}}^{(2)} \right)^2 - Q^2 \right] \left[ Q^2 - \left(
k_{\text{F}}^{(1)} - k_{\text{F}}^{(2)}\right)^2 \right] } } \, .
\end{eqnarray}
Here $\vek{Q} = (e / \hbar)\, d\, \vek{B}_\|\times \vek{\hat z}$ is the magnetic-field-induced
shift in kinetic momentum~\cite{smo89a,eis91,hay91,gen91,sim93,zhe93,lyo93,rai96} for
an electron that has tunneled between two layers spaced vertically at distance $d$.
The factor $g_{\text{s}}=2$ accounts for the real-spin degeneracy, $\rho_{\text{F}}^{(m)}$
is the density of states at the Fermi energy in system $m$ not including real-spin or valley
degrees of freedom, $k_{\text{F}}^{(m)}$ is the Fermi wave vector in system $m$, and
\begin{equation}\label{eq:overlap}
\Gamma_{\text{u/l}}^{(\gamma)}(\vek{Q}) = \, _{\gamma, \vek{k}_{\text{u/l}}-\vek{Q}/2}
\!\! \bra{\sigma^{(1)}_{\text{F}}} \tau_{\vek{k}_{\text{u/l}}}
\ket{\sigma^{(2)}_{\text{F}}}_{\!\gamma, \vek{k}_{\text{u/l}}+\vek{Q}/2}
\end{equation}
are overlap matrix elements between (pseudo-)spinors associated with the electron states
at the two intersection points (labelled u and l, respectively) of  the shifted Fermi circles.
See Fig.~\ref{fig:pseudoSchem} for an illustration.

In the following, we neglect the $\vek{k}$ dependence of the tunneling
matrix~\footnote{This is based on the realistic assumption that the tunnel-barrier height
is much larger than the bandwidth of relevant electronic excitations. In principle, the
$\vek{k}$-dependence of $\tau_{\vek{k}}$ can modify the density dependence
of tunneling conductances but will not affect the functionality of our proposed
valley-filter device.} $\tau_\vek{k}\equiv\tau$ and use the parameterization
\begin{equation}
\tau = \left( \tau_0\,\sigma_0+\tau_x\,\sigma_x+\tau_y \,\sigma_y+\tau_z
\,\sigma_z \right)/\sqrt{2} \quad ,
\end{equation}
with complex numbers $\tau_j$ that encode all possible tunneling processes, including
those that are associated with a (pseudo-)spin flip. To be specific, we limit ourselves
to the case where both the layers are \textit{n}-doped, i.e., where $\sigma^{(1)}_{\text{F}}
=\sigma^{(2)}_{\text{F}}\equiv +$. Because of the rigid locking between the pseudo-spin
state and the kinetic momentum of single-electron eigenstates in single-layer and
bi-layer graphene [see Figs.~\ref{fig:basics}(d)--(f)], it is possible to express the spinors for
positive-energy eigenstates in terms of a rotation matrix $\mathcal{U}(\theta)=\exp(-i \theta
\sigma_z/2)$ and the eigenstates $\ket{\rightarrow}$, $\ket{\leftarrow}$ of pseudo-spin
projection parallel to the $x$ axis as
\begin{subequations}\label{eq:pseudoRot}
\begin{eqnarray}
\ket{\sigma^{(\mathrm{sl})}_{\text{F}}}_{\!\mathbf{K}, \vek{\bar k}}&=&\mathcal{U} \big(
\theta_{\bar k} \big) \ket{\rightarrow}\,\, , \\
\ket{\sigma^{(\mathrm{sl})}_{\text{F}}}_{\!\mathbf{K'}, \vek{\bar k}}&=&\mathcal{U} \big(
\pi - \theta_{\bar k} \big) \ket{\rightarrow}\equiv \sigma_y\, \mathcal{U} \big(
\theta_{\bar k} \big) \ket{\rightarrow}\,\, , \\
\ket{\sigma^{(\mathrm{bl})}_{\text{F}}}_{\!\mathbf{K}, \vek{\bar k}}&=&\mathcal{U} \big(
-2 \theta_{\bar k} \big) \ket{\leftarrow} \,\, , \\
\ket{\sigma^{(\mathrm{bl})}_{\text{F}}}_{\!\mathbf{K'}, \vek{\bar k}}&=&\mathcal{U} \big(
2 \theta_{\bar k} \big) \ket{\leftarrow} \equiv \sigma_x\, \mathcal{U} \big( - 2 \theta_{\bar k}
\big) \ket{\leftarrow} \,\, .
\end{eqnarray}
\end{subequations}
Here we indicated states for electrons in single-layer (bi-layer) graphene by the
superscript (sl) [(bl)], and $\theta_{\bar k} = \arctan(\bar k_y/\bar k_x)$. By virtue of the
matrix element (\ref{eq:overlap}), the magneto-tunneling conductance between graphene
sheets is strongly affected by the spinor structure of electron eigenstates~\cite{vas13,pra13}
and also depends on the pseudo-spin structure of the tunnel barrier~\cite{pra13}.

%\textit{Valley-polarized tunneling transport} -- 
The total current for tunneling through the barrier will be the sum of contributions from
both valleys. However, as we will see below, these contributions need not have equal
weight. To quantify the distribution of tunneling transport between the valleys, we consider
the valley polarization of the conductance defined as
\begin{equation}
\chi =\frac{G^{(\vek{K})} - G^{(\vek{K'})}}{G^{(\vek{K})} + G^{(\vek{K'})}} \quad .
\end{equation}
From Eq.~(\ref{eq:conductance}), we find that $\chi$ is only a function of the pseudo-spin
matrix elements $\Gamma_{\text{u/l}}^{(\gamma)}(\vek{Q})$;
\begin{equation}
\chi =\frac{\left|\Gamma_{\text{u}}^{(\vek{K})}\right|^2 - \left|\Gamma_{\text{u}}^{(\vek{K'})}
\right|^2 + \left| \Gamma_{\text{l}}^{(\vek{K})}\right|^2 - \left|\Gamma_{\text{l}}^{(\vek{K'})}
\right|^2}{\left|\Gamma_{\text{u}}^{(\vek{K})} \right|^2 + \left|\Gamma_{\text{l}}^{(\vek{K})}
\right|^2 + \left| \Gamma_{\text{u}}^{(\vek{K'})} \right|^2 + \left|\Gamma_{\text{l}}^{(\vek{K'})}
\right|^2} \quad .
\end{equation}

\begin{figure}[t]
\includegraphics[width=0.4\columnwidth]{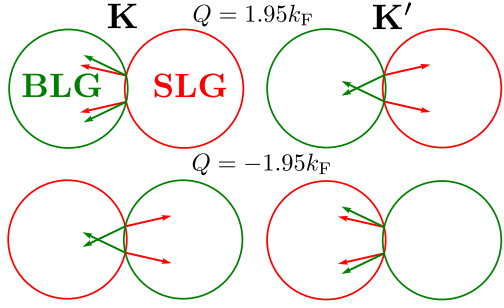}
\caption{\label{fig:fermicircles}
Comparison of pseudo-spin alignment of states from the $\vek{K}$ and
$\vek{K'}$ valleys of single-layer graphene (SLG) and bi-layer graphene
(BLG) at intersection points of their Fermi circles. The case with equal
densities in both layers ($k_{\text{F}}^{({\rm sl})} = k_{\text{F}}^{({\rm bl})}
\equiv k_{\text{F}}$) and $\vek{Q} = Q\, \vek{\hat x}$ with $Q\approx 2
k_\mathrm{F}$ ($Q\approx -2k_\mathrm{F}$) is depicted in the upper
(lower) panel. Note the parallel (opposite) alignment of pseudo-spins for
states from the $\vek{K}$ ($\vek{K'}$) valley as $Q \to 2 k_{\text{F}}$,
and the reversal of alignments for $Q \to -2 k_{\text{F}}$.}
\end{figure}

Without loss of generality, we now consider the situation where the magnetic field
$\vek{B}_\|$ is applied in $\vek{\hat y}$ direction, i.e., $\vek{Q} \equiv Q\,
\vek{\hat x}$. Recognizing the fact that $\vek{k}_\mathrm{u}$ and $\vek{k}_\mathrm{l}$
are then related by mirror symmetry with respect to the $x$ axis allows us to
write the tunnelling matrix elements as
\begin{subequations}\label{eq:matRot}
\begin{eqnarray}
\Gamma_{\text{u}}^{(\vek{K})} &=& \bra{\leftarrow } \mathcal{U} \big(
2\theta_{\vek{k}_{\text{u}}-\vek{Q}/2}\big) \,\, \tau\,\, \mathcal{U} \big(
\theta_{\vek{k}_{\text{u}}+\vek{Q}/2} \big)  \ket{\rightarrow} \,\, ,\\
\Gamma_{\text{l}}^{(\vek{K})} &=& \bra{\leftarrow } \mathcal{U} \big(
2\theta_{\vek{k}_{\text{u}}-\vek{Q}/2}\big) \,\, \sigma_x\, \tau\, \sigma_x \,\, \mathcal{U}
\big( \theta_{\vek{k}_{\text{u}}+\vek{Q}/2} \big)  \ket{\rightarrow} , \quad \\
\Gamma_{\text{u}}^{(\vek{K'})} &=& \bra{\leftarrow } \mathcal{U} \big(
2\theta_{\vek{k}_{\text{u}}-\vek{Q}/2}\big) \,\, \sigma_x\, \tau \, \sigma_y \,\, \mathcal{U}
\big( \theta_{\vek{k}_{\text{u}}+ \vek{Q}/2} \big)  \ket{\rightarrow} ,\\
\Gamma_{\text{l}}^{(\vek{K'})} &=& \bra{\leftarrow } \mathcal{U} \big(
2\theta_{\vek{k}_{\text{u}}-\vek{Q}/2}\big) \,\, \tau\, (-i \sigma_z)\,\, \mathcal{U}
\big( \theta_{\vek{k}_{\text{u}}+\vek{Q}/2} \big)  \ket{\rightarrow} . \quad 
\end{eqnarray}
\end{subequations}
Analysis of the expressions (\ref{eq:matRot}) yields analytical results for the
valley-polarization $\chi$ of the conductance. For example, when $Q = \pm
\big(k_{\text{F}}^{(\mathrm{sl})} + k_{\text{F}}^{(\mathrm{bl})}\big)$, we have
$\theta_{\vek{k}_{\text{u}}\mp\vek{Q}/2} = 0$ and $\theta_{\vek{k}_{\text{u}}
\pm\vek{Q}/2}=\pi$, which yields
\begin{equation}\label{eq:valPolKiss}
\chi\big(\vek{Q} \equiv \pm \big[ k_{\text{F}}^{({\rm sl})} + k_{\text{F}}^{({\rm bl})}
\big] \vek{\hat x}\big) = \pm\frac{|\tau_0 - \tau_x|^2 - |\tau_y + i\tau_z|^2}{|\tau_0
- \tau_x|^2 + |\tau_y + i\tau_z |^2} \quad .
\end{equation}
Thus a nonvanishing $\chi$ is possible depending on the pseudo-spin structure
of the tunneling matrix $\tau$, especially also for the case when pseudo-spin is
conserved in a tunneling event ($\tau \equiv \tau_0 \, \sigma_0/\sqrt{2}$). That
this must be the case can be explained based on the form of pseudo-spin states
from the single-layer and the bi-layer-graphene system for which tunneling is allowed
by simultaneous energy and momentum conservation. See Fig.~\ref{fig:fermicircles}.
For example, as $Q \to k_{\text{F}}^{(\mathrm{sl})} + k_{\text{F}}^{(\mathrm{bl})}$,
the pseudo-spins of states at the `kissing' point of the two Fermi circles become
(oppositely) aligned in the $\vek{K}$ ($\vek{K'}$) valley. Thus if the tunneling
matrix is of the form $\tau\propto (\tau_0\, \sigma_0 + \tau_x\, \sigma_x)$, the
overlap matrix elements (\ref{eq:overlap}) restrict tunneling to occur only for
electrons from the $\vek{K}$ valley. Conversely, if $\tau\propto (\tau_y\, \sigma_y
+ \tau_z\, \sigma_z)$, pseudo-spin must be flipped in a tunneling event, and only
electrons from the $\vek{K'}$ valley are able to accommodate that condition with
simultaneous energy and momentum conservation. In both situations, the tunneling
current will be fully valley-polarized as a result.

\begin{figure}[t]
\includegraphics[width=0.45\columnwidth]{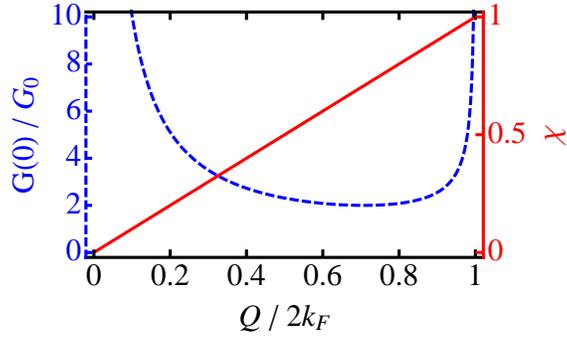}
\caption{\label{fig:polPlot}
Valley polarization $\chi$ of the tunneling conductance (red solid curve) and total
tunneling conductance $G$ (blue dashed curve) for pseudo-spin-conserving
tunneling between vertically separated sheets of single-layer and bi-layer graphene.
having equal electron densities. The conductance unit is $G_0= e^2 M A\,
|\tau_0|^2/(2\pi \hbar^4 v k_\mathrm{F})$.}
\end{figure}

The tunneling current will be non-vanishing in general and can even be large under
the same conditions that maximize $\chi$. As an example, we provide the expression
of the \emph{total\/} tunneling conductance for the case when $\tau=\left(\tau_0
\sigma_0 + \tau_x\sigma_x \right)/\sqrt{2}$:
\begin{equation}\label{eq:totCond}
G \equiv G^{(\vek{K})} + G^{(\vek{K'})} = \frac{e^2}{\pi\hbar} \, \frac{M\,
k_\mathrm{F}^{(\mathrm{sl})}A}{v \hbar^3} \frac{2 | \tau_0 + \tau_x |^2 - {\rm Re}
\{ \tau_0\, \tau_x^\ast \} \left[ \frac{ Q^2 + \left( k_\mathrm{F}^{(\mathrm{bl})}\right)^2
- \left( k_\mathrm{F}^{(\mathrm{sl})}\right)^2}{Q \, k_\mathrm{F}^{(\mathrm{bl})}}
\right]^2}{\sqrt{\left[ \left( k_\mathrm{F}^{(\mathrm{bl})} +
k_\mathrm{F}^{(\mathrm{sl})} \right)^2 - Q^2 \right] \left[ Q^2 - \left(
k_\mathrm{F}^{(\mathrm{bl})} - k_\mathrm{F}^{(\mathrm{sl})} \right)^2 \right]}}
\, , 
\end{equation}
where $M$ is the effective mass of electrons in the graphene bi-layer~\cite{cas09}, $v$
the Fermi velocity in the single layer~\cite{cas09}, and $A$ the area of the tunnel-barrier
interface between the vertically separated 2D conductors. Clearly, for the condition $Q=
\pm \big[ k_{\text{F}}^{(\mathrm{sl})} + k_{\text{F}}^{(\mathrm{bl})}\big]$ associated with
100\% valley polarization of the tunneling current, there is a divergence in the total
magneto-tunnelling conductance.~\footnote{In real samples, the finite electronic life time
will smoothen this divergence into a Lorentzian peak. See, e.g., Ref.~\onlinecite{jun96}.}
Specializing $\tau = \tau_0\, \sigma_0/\sqrt{2}$, we can give the result
\begin{equation}
\chi(\vek{Q}) = - \cos \left( \theta_{\vek{k}_{\text{u}}+\vek{Q}/2} + 2\theta_{\vek{k}_{\text{u}}
-\vek{Q}/2} \right) \quad ,
\end{equation}
which further simplifies when $k_{\text{F}}^{(\mathrm{sl})} =
k_{\text{F}}^{(\mathrm{bl})}\equiv k_\mathrm{F}$ to 
%\begin{equation}
$\chi (\vek{Q}) = Q/2 k_\mathrm{F}$.
%\end{equation}
See Fig.~\ref{fig:polPlot} for an illustration of the simultaneous occurrence of 100\% valley
polarization and maximum of tunneling transport, shown there for the special situation of
equal densities in the two layers and a pseudo-spin-conserving tunnel barrier.

%\textit{Experimental requirements} -- 
The efficiency of the valley-filtering device proposed here will be affected by the pseudo-spin
structure of the tunnel barrier, which is determined by the geometric placement of the single
and bi-layer sheets with respect to each other. Our previously suggested method~\cite{pra13}
to determine the full pseudo-spin structure of the tunnel coupling could be employed to
optimize the vertical-heterostructure design in this regard. Furthermore, the valley polarization
of the tunneling conductance is limited by the available magnitudes of the in-plane magnetic
field. Using the case of equal density in the two layers and pseudo-spin-conserving tunneling
as an example, we can estimate
\begin{equation}
\chi \le \min\left\{1, \frac{e }{\hbar}\, \frac{B^{(\mathrm{max})} d}{\sqrt{4\pi n}} \equiv 0.05 \times
\frac{B^{(\mathrm{max})}\,[\mathrm{T}]\,\, d\, [\mathrm{nm}]}{\sqrt{n\,[10^{10} \, \mathrm{cm}^{-2}]}}
\right\} \, .
\end{equation}
Thus in-plane magnetic fields of the order of $10$~T are required to generate significant valley
polarization in realistic vertical heterostructures of graphene layers.

%\textit{Conclusions} --
In conclusion, we have studied tunneling transport between two parallel graphene sheets, one
being a single-layer and the other a bi-layer sample. The requirement of simultaneous energy and
momentum conservation, together with the distinctive valley-contrasting pseudo-spin-momentum
locking in the two different graphene systems, causes a finite valley polarization of the tunneling
current when an in-plane magnetic field is applied. For large-enough field magnitude, 100\% valley
polarization can be achieved, and a significant magnitude of  polarization is generally realized
concomitantly with large values of the total tunneling current.

In contrast to many other valley-filter designs, the vertical-tunneling-based proposal works
without substantially altering the conducting properties of the graphene sheets. As a
valley-polarized current is generated in the  bulk, the present set-up is ideal for realizing
valley-optoelectronic devices~\cite{yao08,gol11} as well as applications related to the valley-Hall
effect~\cite{xia07} and its inverse. The fact that a valley polarization
simultaneously exists in parallel single and bi-layer graphene sheets opens up possibilities
for a three-dimensional valleytronic chip design.

\textit{Acknowledgments} --
LP gratefully acknowledges scholarship funding from Victoria University. Work at KITP
was supported in part by the National Science Foundation under Grant No.\ NSF PHY11-25915.

%\bibliography{MagTunValFilt}

\end{document}